\documentclass[twocolumn]{aastex63}

\usepackage[fleqn]{amsmath}
\usepackage[hyphenbreaks]{breakurl}
\usepackage{float}
\usepackage{url}

\accepted{Feb 9, 2023}
\submitjournal{ApJL}

\shorttitle{NIR Period-Luminosity-Metallicity relations for RR Lyrae variables}
\shortauthors{Bhardwaj A. et al.}

\begin{document}

\title{Precise Empirical Determination of Metallicity Dependence of Near-infrared Period-Luminosity Relations for RR Lyrae Variables}

\correspondingauthor{Anupam Bhardwaj}
\email{anupam.bhardwajj@gmail.com; anupam.bhardwaj@inaf.it}
\author[0000-0001-6147-3360]{Anupam Bhardwaj}\thanks{Marie-Curie Fellow}
\affil{INAF-Osservatorio Astronomico di Capodimonte, Via Moiariello 16, 80131 Napoli, Italy}
\author[0000-0002-1330-2927]{Marcella Marconi}
\affil{INAF-Osservatorio Astronomico di Capodimonte, Via Moiariello 16, 80131 Napoli, Italy}
\author[0000-0002-6577-2787]{Marina Rejkuba}
\affiliation{European Southern Observatory, Karl-Schwarzschild-Stra\ss e 2, 85748, Garching, Germany}
\author[0000-0002-7203-5996]{Richard de Grijs}
\affil{School of Mathematical and Physical Sciences, Macquarie University, Balaclava Road, Sydney NSW 2109, Australia}
\affil{Research Centre for Astronomy, Astrophysics and Astrophotonics, Macquarie University, Balaclava Road, Sydney NSW 2109, Australia}
\author[0000-0001-6802-6539]{Harinder P. Singh}
\affiliation{Department of Physics and Astrophysics, University of Delhi,  Delhi-110007, India }
\author{Vittorio F. Braga}
\affil{INAF-Osservatorio Astronomico di Roma, via Frascati 33, 00040 Monte Porzio Catone, Italy}
\author[0000-0003-0791-7594]{Shashi Kanbur}
\affiliation{Department of Physics, State University of New York, Oswego, NY 13126, USA}
\author[0000-0001-8771-7554]{Chow-Choong Ngeow}
\affil{Graduate Institute of Astronomy, National Central University, 300 Jhongda Road, 32001 Jhongli, Taiwan}
\author{Vincenzo Ripepi}
\affil{INAF-Osservatorio Astronomico di Capodimonte, Via Moiariello 16, 80131 Napoli, Italy}
\author{Giuseppe Bono}
\affil{INAF-Osservatorio Astronomico di Roma, via Frascati 33, 00040 Monte Porzio Catone, Italy}
\affil{Dipartimento di Fisica, Università di Roma Tor Vergata, via della Ricerca Scientifica 1, 00133 Roma, Italy}
\author[0000-0002-5819-3461 
]{Giulia De Somma}
\affil{INAF-Osservatorio Astronomico di Capodimonte, Via Moiariello 16, 80131 Napoli, Italy}
\affil{INFN-Sez. di Napoli, Compl. Univ. di Monte S. Angelo, Edificio G, Via Cinthia, 80126 Napoli, Italy}
\author{Massimo Dall'Ora}
\affil{INAF-Osservatorio Astronomico di Capodimonte, Via Moiariello 16, 80131 Napoli, Italy}

\begin{abstract} 
RR Lyrae variables are excellent population II distance indicators thanks to their well-defined period-luminosity relations (PLRs) at infrared wavelengths. We present results of near-infrared (NIR) monitoring of Galactic globular clusters to empirically quantify the metallicity dependence of NIR PLRs for RR Lyrae variables. Our sample includes homogeneous, accurate, and precise photometric data for 964 RR Lyrae variables in 11 globular clusters covering a large metallicity range ($\Delta\textrm{[Fe/H]}\sim2$~dex). We derive $JHK_s$ band period-luminosity-metallicity (PLZ) and period-Wesenheit-metallicity (PWZ) relations anchored using 346 Milky Way field RR Lyrae stars with {\it Gaia} parallaxes, and simultaneously solved for independent distances to globular clusters. We find a significant metallicity dependence of $\sim0.2$~mag/dex in $JHK_s$ band PLZ and PWZ relations for RR Lyrae stars independent of the adopted metallicity scale. The metallicity coefficients and the zero-points of the empirical PLZ and PWZ relations are in excellent agreement with the predictions from the horizontal branch evolution and pulsation models. Furthermore, RR Lyrae based distances to our sample of globular clusters are also statistically consistent with other independent measurements in the literature. Our recommended empirical $JHK_s$ band PLZ relations for RR Lyrae stars with periods of fundamental mode pulsation ($P_\textrm{f}$) are:
\begin{eqnarray}
   ~~~~~~~~M_J =-0.44~(\pm0.03) -1.83~(\pm0.02)\log(P_\textrm{f}) + 0.20~ (\pm0.02)~\textrm{[Fe/H]}~~(\sigma=0.05~\textrm{mag}) \nonumber \\ 
  ~~~~~~~~M_H =-0.74~(\pm0.02) -2.29~(\pm0.02)\log(P_\textrm{f}) + 0.19~ (\pm0.01)~\textrm{[Fe/H]}~~(\sigma=0.05~\textrm{mag}) \nonumber \\ 
  ~~~~~~~~M_{K_s} =-0.80~(\pm0.02) -2.37~(\pm0.02)\log(P_\textrm{f}) + 0.18~ (\pm0.01)~\textrm{[Fe/H]}~~(\sigma=0.05~\textrm{mag}) \nonumber  \\ \nonumber
\end{eqnarray}

\end{abstract} 

\section{Introduction}

RR Lyrae variables are old ($> 10$ Gyr), low-mass ($\sim0.5-0.8 M_\odot$) stars that are located at the intersection of the horizontal branch and the classical instability strip in the Hertzsprung-Russell diagram. These pulsating stars have long been used as distance indicators thanks to the correlation of their visual magnitudes with metallicity \citep[see the reviews by][and references therein]{sandage2006, catelan2015, beaton2018, bhardwaj2020}. Since the pioneering work by \citet{longmore1986} it has been known that RR Lyrae stars follow a well-defined period-luminosity relation (PLR) in $K_s$-band. Near-infrared (NIR) PLRs offer advantages over visual magnitude-metallicity relation for distance determinations due to smaller dependence on interstellar extinction, smaller variability amplitude, and less sensitivity to evolutionary effects \citep{bono2001, catelan2004, marconi2015}.  

Stellar evolution and pulsation models of RR Lyrae stars have predicted a strong effect of metallicity on NIR PLRs \citep[$\sim0.18$~mag/dex in $JHK_s$,][]{catelan2004, marconi2015}. The empirical studies however have reported a wide range for the metallicity coefficient of period-luminosity-metallicity (PLZ) relations \citep[$0.03-0.23$~mag/dex in $K_s$,][and references therein]{ sollima2006, sollima2008, bono2011,muraveva2015,muraveva2018, braga2018,layden2019, neeley2019, muhie2021, cusano2021}. Most of these empirical investigations were carried out only in $K_s$-band, and very few of them included all three  $JHK_s$ filters. \citet{neeley2019} used Milky Way  (MW) field RR Lyrae stars with parallaxes from {\it Gaia} data release 2 \citep{prusti2016, brown2018} to derive PLZ relations in NIR bands. These authors found a metallicity coefficient which is consistent with theoretical predictions but with a four times larger scatter. \citet{bhardwaj2021} showed that the lack of homogeneous NIR photometry and spectroscopic metallicities limits the precision of PLZ relation for field RR Lyrae stars in the MW despite the improvements in their parallax measurements with the early {\it Gaia} data release 3 \citep{brown2021}. The latest {\it Gaia} parallaxes of MW field RR Lyrae stars have also been used to calibrate PLZ relations at mid-infrared wavelengths based on Wide-field Infrared Survey Explorer data \citep[e.g.,][]{gilligan2021, mullen2023}.

RR Lyrae variables in globular clusters (GCs) of different mean metallicities provide a unique opportunity to quantify the metallicity coefficient of PLZ relations. A previous such study was carried out by \citet{sollima2006} in $K_s$-band using literature photometry of 538 RR Lyrae stars in 16 GCs with $\Delta \textrm{[Fe/H]}=1.3$~dex, which resulted in a weak constraint on the metallicity coefficient ($0.08\pm0.11$ mag/dex). This work aims to use homogeneous $JHK_s$ time-series photometry of RR Lyrae variables in GCs from our ongoing observational program to empirically calibrate their PLZ relations and extinction-free period-Wesenheit-metallicity (PWZ) relations. The empirical calibrations of RR Lyrae PLZ and PWZ relations at NIR wavelengths will be crucial for stellar astrophysics and distance scale studies based on population II standard candles. The well-established infrared PLZ relations for RR Lyrae stars will be particularly important for the James Webb Space Telescope, upcoming Nancy Grace Roman Space Telescope, and ground-based thirty meter class telescopes, which will operate predominantly at infrared wavelengths and enable observations of a large number of RR Lyrae stars in galaxies beyond the Local Group.

\section{The Data}
\label{sec:data}

\begin{figure}
\centering
\includegraphics[width=0.42\textwidth]{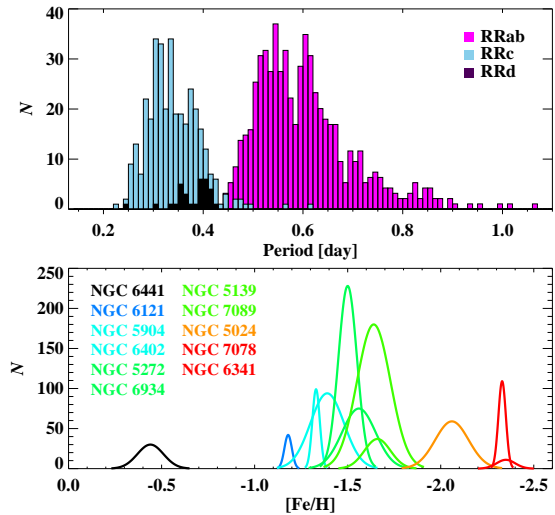}\\
\caption{Histograms of period distributions (top) of 964 GC RR Lyrae variables and the distribution of mean metallicities (bottom) for their parent GCs. The peak and the width of the metallicity distribution represent the mean metallicity and its scatter from \citet{carretta2009}, respectively.} 
\label{fig:hists}
\end{figure}

\begin{deluxetable*}{rcccccccc}
\huge
\tablecaption{Globular clusters with $JHK_s$ time-series photometric data. \label{tbl:gc_data}}
\tabletypesize{\footnotesize}
\tablewidth{0pt}
\tablehead{\colhead{\bf GC} &   &  \multicolumn{2}{c}{\bf $N$} & \colhead{\bf Ref.$^a$}  & \multicolumn{1}{c}{\bf [Fe/H]$^b$} & \multicolumn{1}{c}{\bf E(B-V)$^c$} & \colhead{\bf $\mu_\textrm{GC}^d$} & \colhead{\bf $\mu_\textrm{TW}^e$}\\
        &   & RRab & RRc &    &  dex &  mag & mag & mag }
\startdata
NGC~6441&             -&   21&    9& B22& $   -0.44\pm0.07    $ &    0.47&    $15.52\pm0.03$    &    $15.48\pm0.02$    \\
NGC~6121&            M4&   30&   12& S14& $   -1.18\pm0.02    $ &    0.35&    $11.34\pm0.02$    &    $11.29\pm0.02$    \\
NGC~5904&            M5&   67&   32& B23& $   -1.33\pm0.02    $ &    0.03&    $14.37\pm0.02$    &    $14.30\pm0.02$    \\
NGC~6402&           M14&   47&   47& B23& $   -1.39\pm0.09    $ &    0.60&    $14.80\pm0.06$    &    $14.75\pm0.02$    \\
NGC~5272&            M3&  171&   57& B20& $   -1.50\pm0.05    $ &    0.01&    $15.04\pm0.02$    &    $15.02\pm0.02$    \\
NGC~6934&             -&   63&   12& B23& $   -1.56\pm0.09    $ &    0.10&    $15.98\pm0.02$    &    $15.95\pm0.02$    \\
NGC~5139&  $\omega$~Cen&   83&   97& B18& $   -1.64\pm0.09    $ &    0.12&    $13.67\pm0.02$    &    $13.67\pm0.02$    \\
NGC~7089&            M2&   22&   15& B23& $   -1.66\pm0.07    $ &    0.06&    $15.34\pm0.02$    &    $15.31\pm0.03$    \\
NGC~5024&           M53&   28&   31& B21a& $   -2.06\pm0.09    $ &    0.02&    $16.34\pm0.02$    &   $16.35\pm0.03$    \\
NGC~7078&           M15&   48&   61& B21b& $   -2.33\pm0.02    $ &    0.10&    $15.15\pm0.02$    &   $15.14\pm0.03$    \\
NGC~6341&           M92&    8&    3& D05& $   -2.35\pm0.05    $ &    0.02&    $14.65\pm0.02$    &    $14.65\pm0.04$    \\
\enddata
\tablecomments{$^a$References: B22 - \citet{bhardwaj2022a}, S14 - \citet{stetson2014}, B20 - \citet{bhardwaj2020a}, B18 - \citet{braga2018}, B21a - \citet{bhardwaj2021a}, B21b - \citet{bhardwaj2021}, D05 - \citet{del2005}, B23 - Bhardwaj et al. (in prep.).\\
$^b$The mean metallicities were taken from \citet{carretta2009}.\\
$^c$The reddening values were adopted from \citet{harris2010} with an average uncertainty of 0.02~mag, which represents the scatter in $E(B-V)$ values from \citet{harris2010} and \citet{green2019}.\\
$^d$The homogeneous distance moduli ($\mu_\textrm{GC}$) are from the catalog of \citet{baumgardt2021}. \\
$^e$The average distance moduli to GCs derived in this work ($\mu_\textrm{TW}$) using NIR PLZ and PWZ relations in the form of equation~(\ref{eq:plzr_gc}) for the whole sample of 1310 RR Lyrae stars.}
\vspace{-5pt}
\end{deluxetable*}

We use NIR photometry of RR Lyrae variables in GCs from our ongoing observational programs using the WIRCam instrument at the Canada-France-Hawaii Telescope (CFHT) and the FLAMINGOS-2 instrument at the Gemini South telescope\footnote{ Interested readers are referred to \citet{bhardwaj2020a} and \citet{bhardwaj2022a} for details regarding the observing strategy, the data reduction, and analysis of the data from CFHT and Gemini telescopes, respectively.}. All the photometric data for GCs have been calibrated to Two Micron All Sky Survey \citep[2MASS,][]{skrutskie2006} photometric system. Time-series photometry has been published for NGC 5272, NGC 5024, NGC 7078, and NGC 6441 \citep{bhardwaj2020a, bhardwaj2021, bhardwaj2021a, bhardwaj2022a}, and will be presented for NGC 5904, NGC 6402, NGC 6934, and NGC 7089 in a subsequent study (Bhardwaj et al. in prep.). We also included homogeneous $JHK_s$ mean magnitudes for RR Lyrae variables in NGC 6341 \citep{del2005}, NGC 6121 \citep{stetson2014}, and NGC 5139 \citep{braga2018}. Our sample consists of 964 stars including 588 fundamental mode (RRab), 345 first-overtone  (RRc), and 31 mixed-mode RR Lyrae (RRd) variables. The top panel of Fig.~\ref{fig:hists} shows period distributions of 964 GC RR Lyrae stars and the bottom panel shows the distribution of mean metallicities for their parent GCs. Table~\ref{tbl:gc_data} lists the GCs which host RR Lyrae with $JHK_s$ time series photometric data together with their homogeneous distances, mean metallicities, and reddening values. 

\begin{figure}
\centering
\includegraphics[width=0.42\textwidth]{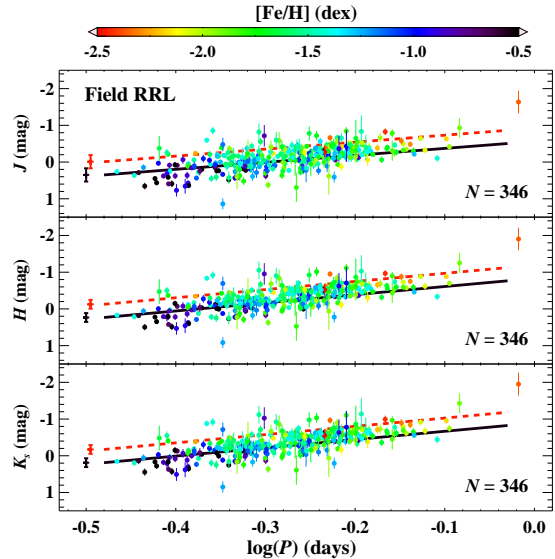}\\
\caption{Near-infrared PLRs for MW field RR Lyrae variables in $J$ (top), $H$ (middle) and $K_s$ (bottom) bands. The solid black and dashed red lines represent theoretical PLZ relations for [Fe/H]$=-0.4$~dex and $-2.4$~dex, respectively. The representative error bars on the left display $\pm 3\sigma$ scatter in the theoretical relations \citep{marconi2015}.} 
\label{fig:calib}
\end{figure}

To calibrate the absolute zero-point of NIR PLZ relations, we also included MW field RR Lyrae stars with {\it Gaia} data and $JHK_s$ photometry from 2MASS. The initial sample of $\sim400$ stars with positions, metallicities, and reddening values from the literature was taken from \citet{muraveva2018}. This sample was cross-matched,  within a $1\arcsec$ search radius, with the 2MASS catalog and the {\it Gaia} distances and RR Lyrae variability catalogs \citep{bailer2021, clementini2022}. The combined sample of 382 stars was further restricted to have metallicity range ($-0.3 < \textrm{[Fe/H]}<-2.5$~dex) similar to that of cluster variables and have 2MASS quality flag `A' in $JHK_s$ filters (346 stars). We did not apply random-phase corrections to single-epoch 2MASS magnitudes because no proper quantification of amplitude ratios or phase lags is available between $G$ and $JHK_s$ bands, and these observations are not contemporaneous. A comparison of intensity-averaged magnitudes of 342 RR Lyrae stars in GCs with single-epoch 2MASS magnitudes (with quality flag `A' in $JHK_s$) results in a median offset of $0.026/0.015/0.051$~mag and a scatter of $0.13/0.11/0.12$~mag in $J/H/K_s$. These systematic offsets were propagated to the photometric uncertainties on 2MASS magnitudes of MW field stars.

Fig.~\ref{fig:calib} displays $JHK_s$ PLRs for MW field RR Lyrae stars using the final sample of 346 field variables (321 RRab and 25 RRc stars). The 2MASS $JHK_s$ magnitudes were corrected for reddening assuming the reddening law of \citet{card1989} with an $R_V=3.23$, and the total-to-selective absorption ratios in NIR filters as: $A_{J/H/K_s}=0.94/0.58/0.39E(B-V)$. The periods and classification of RR Lyrae variables from the {\it Gaia} catalog \citep{clementini2022}, literature reddening estimates \citep{muraveva2018} and the {\it Gaia} based distances \citep{bailer2021} were used to obtain absolute magnitudes for individual stars. A large scatter in the PLZ relations ($0.20/0.17/0.16$ mag in $J/H/K_s$) for MW field stars is seen in Fig.~\ref{fig:calib} as compared to the theoretical relations \citep[$0.06/0.04/0.04$ mag in $J/H/K_s$,][]{marconi2015}, which is not only due to random-phase magnitudes but also due to parallax uncertainties, heterogeneous metallicity and reddening estimates. Nevertheless, the slopes of NIR PLRs for RR Lyrae variables in GCs are well constrained \citep{bhardwaj2022}, which will allow an accurate determination of absolute zero-points of PLZ and PWZ relations at $JHK_s$ wavelengths using MW field stars. This analysis will use the combined sample of 1310 RR Lyrae stars from GCs (964 sources) and MW field (346 sources).

\section{Linear regression formalism}
\label{sec:lfit}

Our goal is to simultaneously solve for a common slope and a metallicity coefficient, and a separate zero-point for each GC. This approach does not require a prior assumption of an independent distance to a given GC, which can be obtained simultaneously given an absolute zero-point based on field RR Lyrae stars. In the mathematical form, the extinction-corrected apparent magnitude or the Wesenheit magnitude of the $j^\textrm{th}$ RR Lyrae star in $i^\textrm{th}$ GC at wavelength $\lambda$ is defined as:

\begin{eqnarray}
    m_{\lambda_{i,j}} = \mu_i + a_\lambda + b_\lambda \log(P_{i,j}) + c_\lambda \textrm{[Fe/H]}_{i,j},
    \label{eq:plzr_gc}
\end{eqnarray}

\noindent where, $P_{i,j}$ is the period in days and [Fe/H]$_{i,j}$ is the metallicity of RR Lyrae stars. 
For a given cluster, the mean metallicity from \citet{carretta2009}, listed in Table~\ref{tbl:gc_data}, was adopted for all RR Lyrae members. The term ($\mu_i+a_\lambda$) also gives the apparent zero-point of the PLZ/PWZ relation in a given cluster. In the case of field variables, the absolute magnitude of a given RR Lyrae in the above equation is:

\begin{eqnarray}
    M_{\lambda_{j}} = a_\lambda + b_\lambda \log(P_{j}) + c_\lambda \textrm{[Fe/H]}_{j}.
    \label{eq:plzr_mw}
\end{eqnarray}

To solve equations~(\ref{eq:plzr_gc}) and (\ref{eq:plzr_mw}) simultaneously, we setup a matrix formalism where vector $Y$ includes all the magnitude measurements. The set of equations are given by a matrix $L$ of $n_Y$ rows and $n_Q$ columns. Here, $n_Y$ is the total number of magnitude measurements in  $Y$ and $n_Q$ is the number of free parameters (11 GC distances, zero-point, slope, and metallicity coefficient) in the vector $Q$. The $\chi^2$ statistics is given as $\chi^2=(Y-LQ)^TC^{-1}(Y-LQ)$, where $C$ is the error matrix. The maximum-likelihood parameters are obtained using $Q=(L^TC^{-1}L)^{-1}L^TC^{-1}Y$, while the standard errors on the free parameters come from the covariance matrix $(L^TC^{-1}L)^{-1}$. This setup of matrix equations is similar to those described in detail in \citet[][in the appendix]{riess2016}. This matrix formalism is employed to obtain free parameters $Q$, which are used to compute residuals of the best-fitting linear-regression. This process is iterated by removing the single largest outlier in each iteration until all residuals are within $3\sigma$, where $\sigma$ is the standard deviation of the global fit.  We created $10^3$ random realizations by varying the magnitudes, adopted mean-metallicity and reddening values within their uncertainties to obtain final coefficients and their associated uncertainties.  

\section{Period-Luminosity-Metallicity relations}

\begin{figure*}
\centering
\gridline{\fig{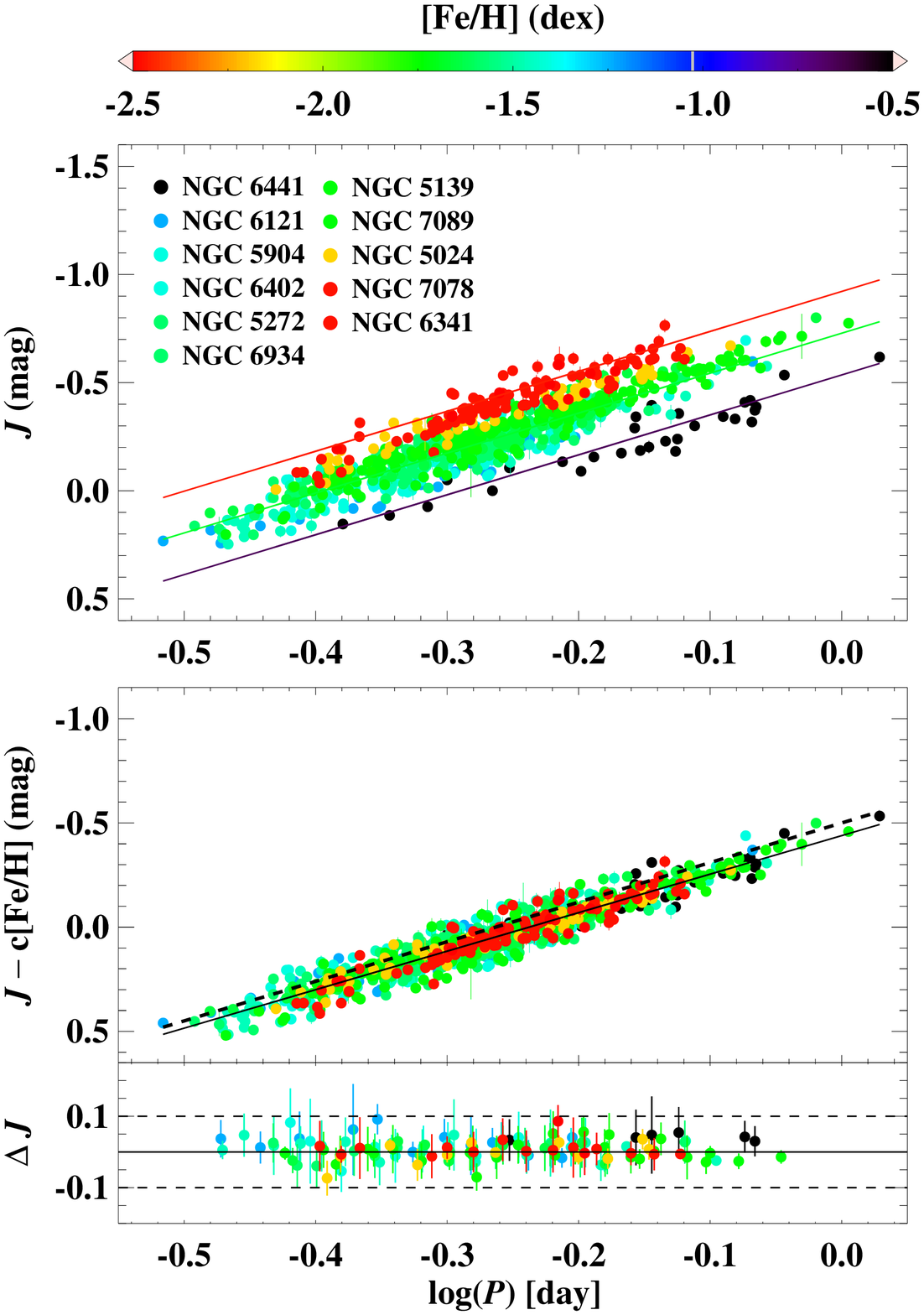}{0.33\textwidth}{}
          \fig{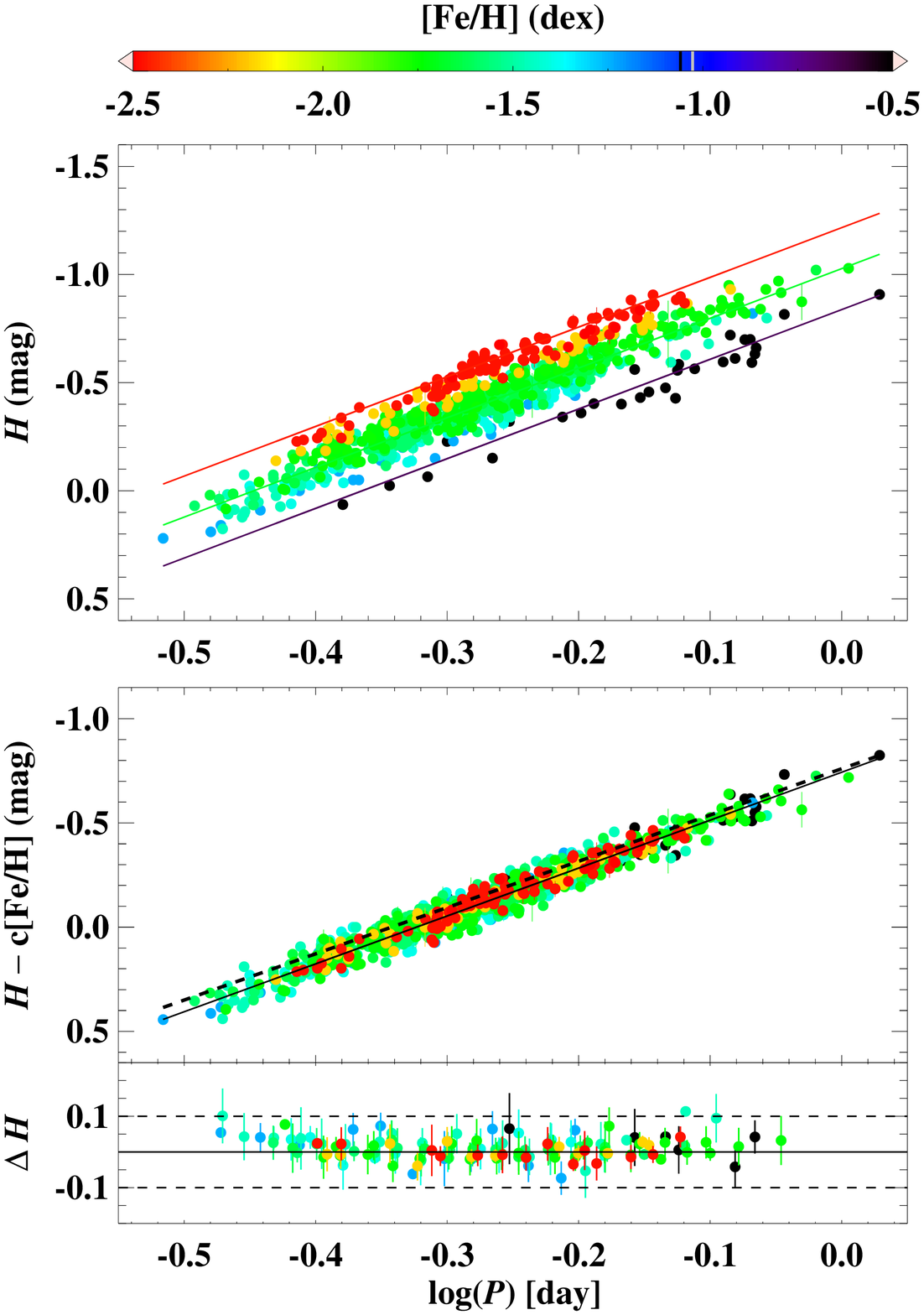}{0.33\textwidth}{}
          \fig{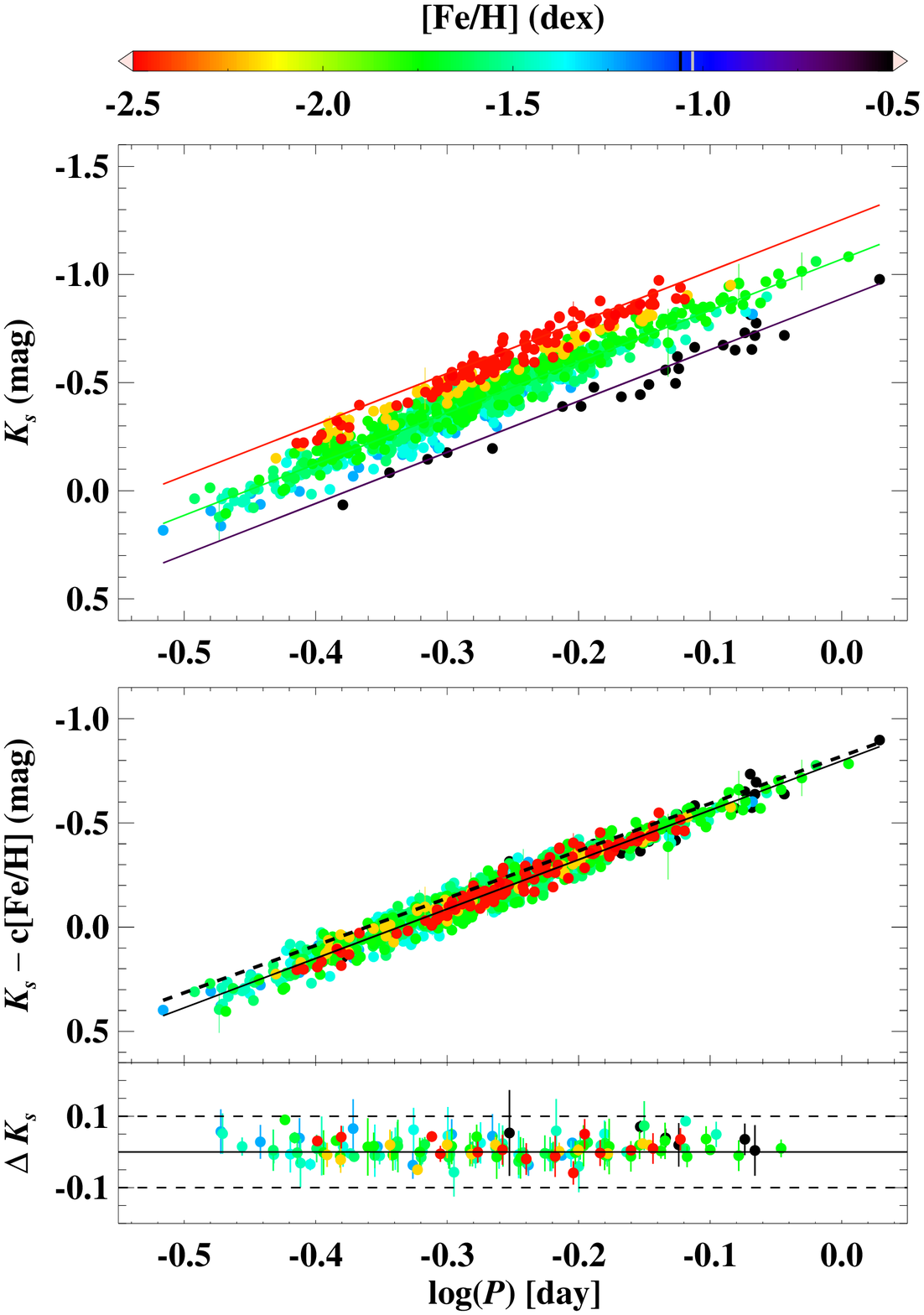}
          {0.33\textwidth}{}}
\vspace{-10pt}
\gridline{\fig{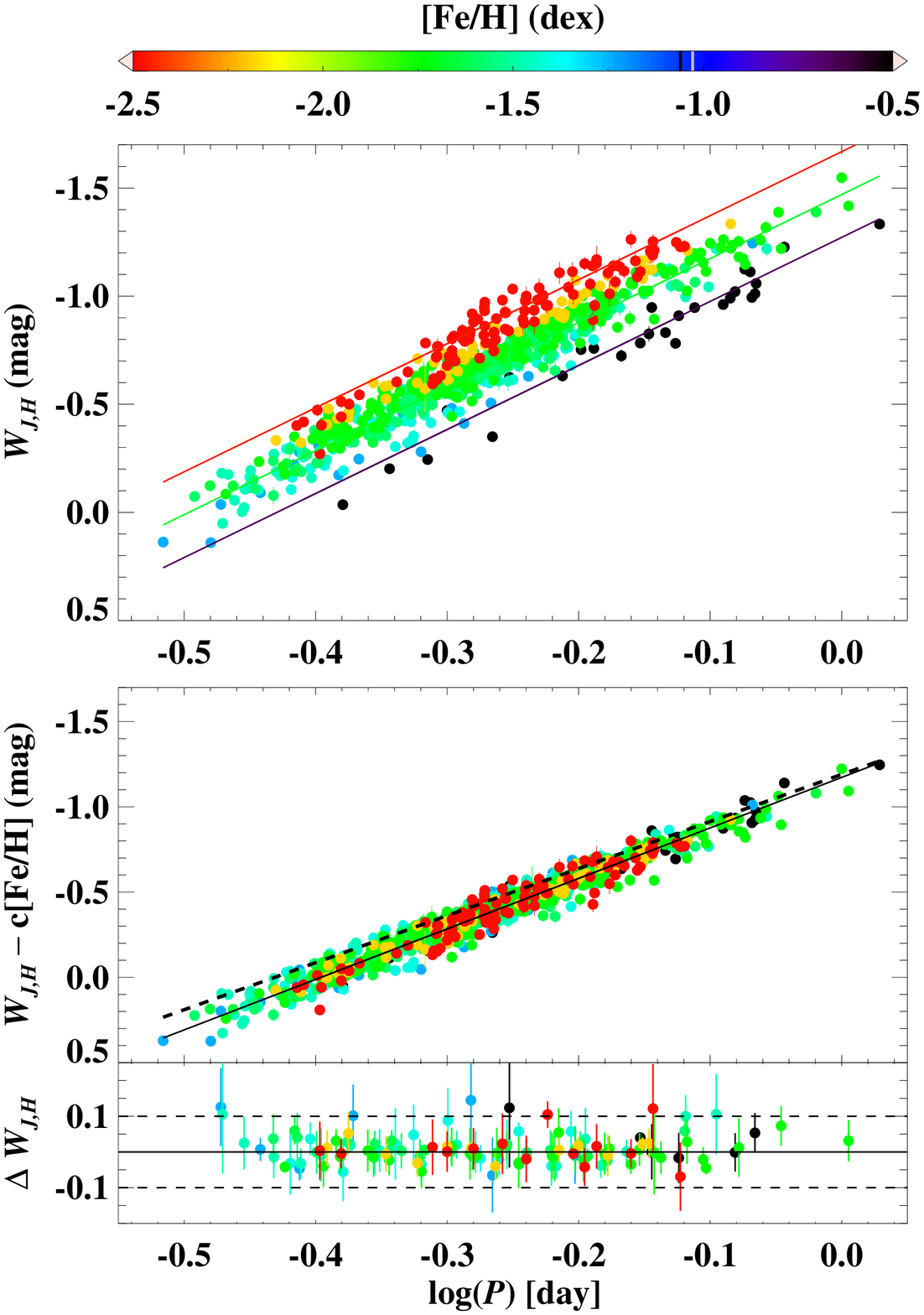}{0.33\textwidth}{}
          \fig{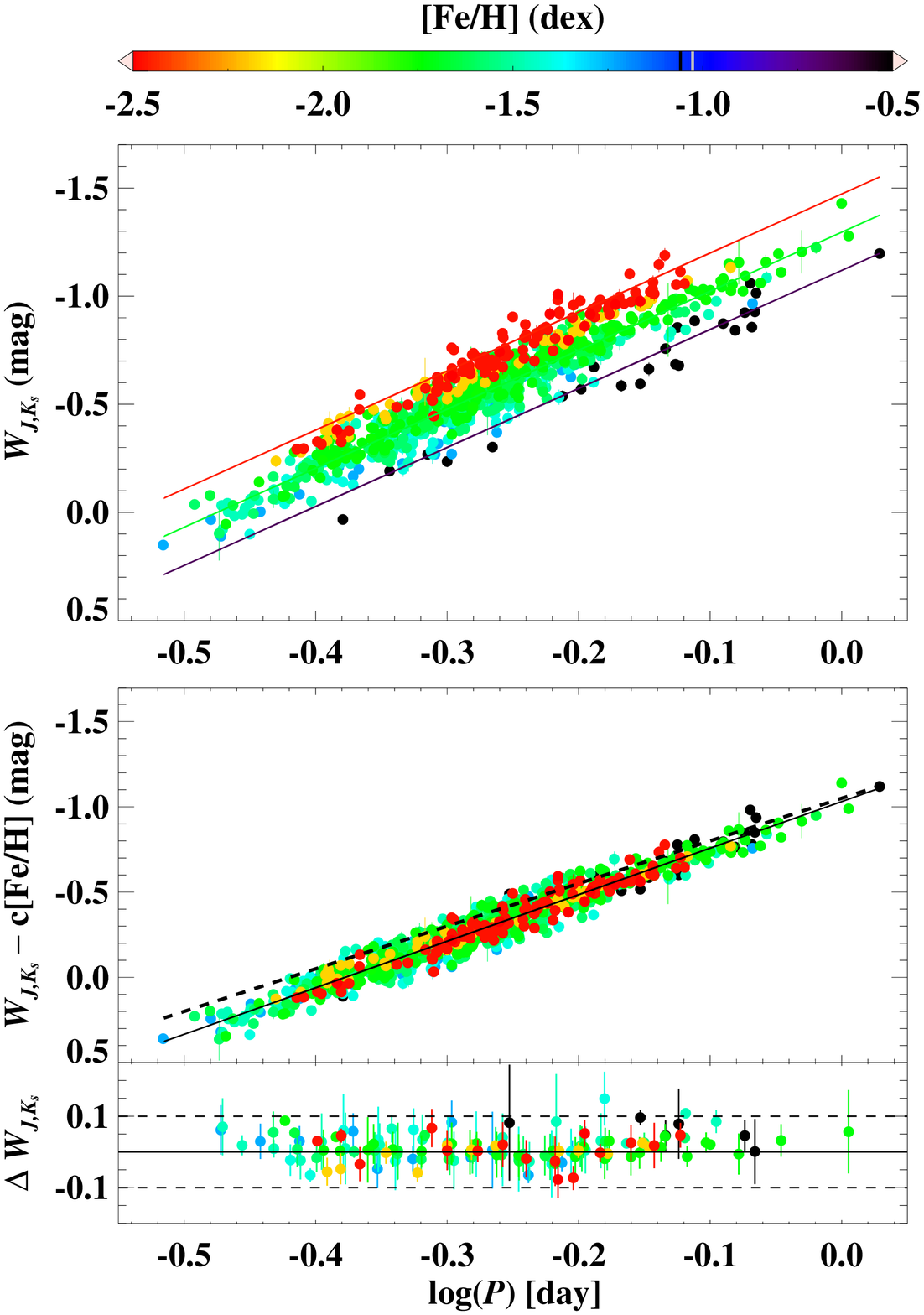}{0.33\textwidth}{}
          \fig{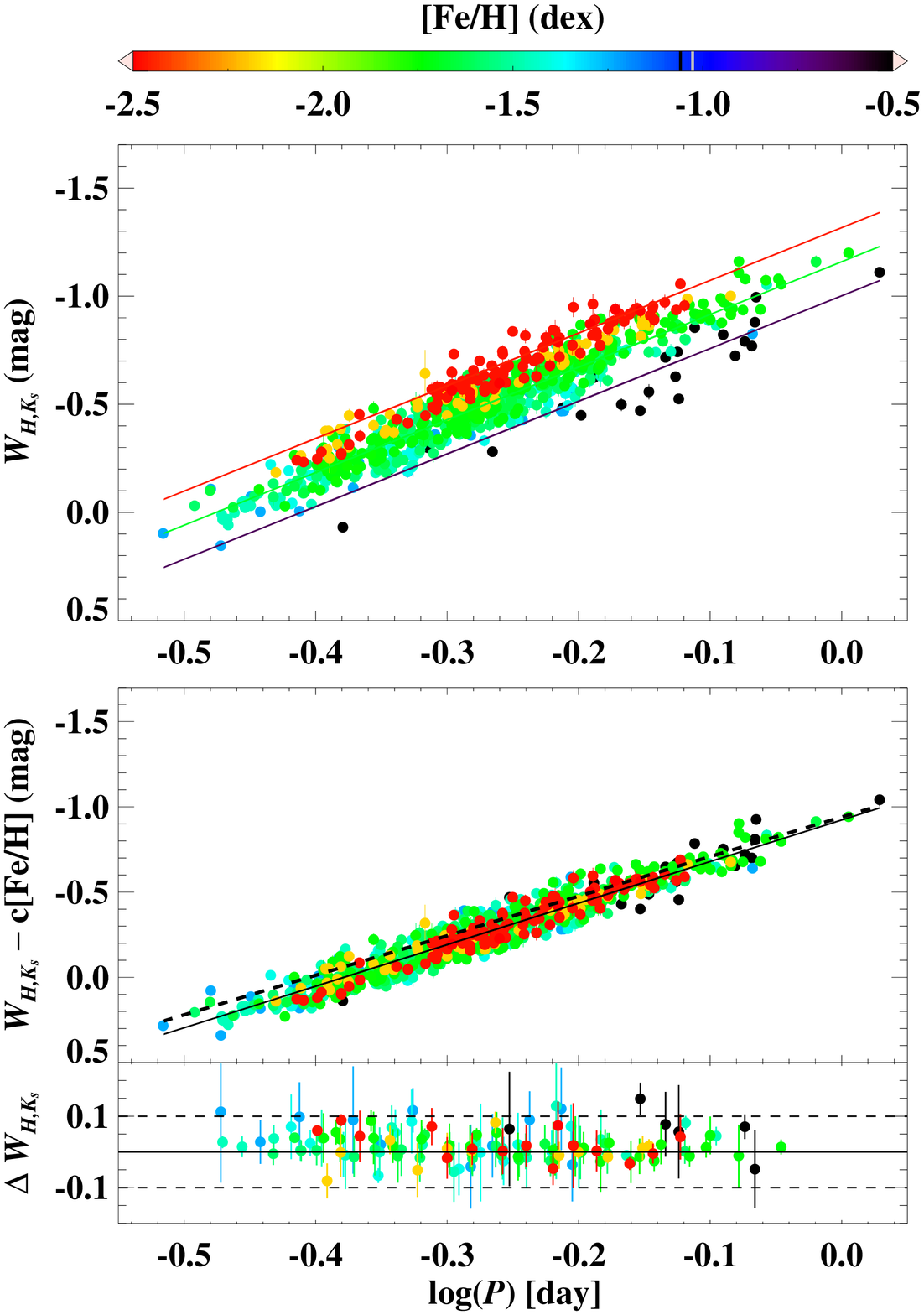}{0.33\textwidth}{}}
\vspace{-10pt}
\caption{Near-infrared PLZ and PWZ relations for RR Lyrae variables in GCs in $JHK_s$ bands. In each case, the color-coded solid lines in the top panels correspond to [Fe/H]=[$-0.5,-1.5,-2.5$]~dex. The solid line in the bottom panels shows the best-fitting linear regression while the dashed line displays the predicted relation from \citet{marconi2015}. The average residuals ($\Delta J,~\Delta H,~\Delta K_s$) in a period bin of $\log(P)=0.02$~days are shown for visualization purposes. The data points are color-coded according to the color bar.} 
\label{fig:plzrs}
\end{figure*}

We used $JHK_s$ mean magnitudes for 964 RR Lyrae variables in GCs and 2MASS single epoch magnitudes for 346 MW field stars jointly to derive the PLZ and PWZ relations. Three different samples of RR Lyrae stars were considered separately: RRab, RRc+RRd, and the combined sample of all 1310 RR Lyrae variables.
 The mixed-mode RRd variables have the dominant first-overtone periods as seen in Fig.~\ref{fig:hists}. Therefore, the first-overtone periods of RRc and RRd stars were fundamentalized using the equation: $\log(P_{f}) = \log(P_{o}) +0.127$ \citep[e.g.,][]{coppola2015, braga2022} for the combined sample of all 1310 RR Lyrae stars. 

The intensity-averaged $JHK_s$ magnitudes for variables in GCs were corrected for reddening assuming the previously mentioned reddening law of \citet{card1989} and absorption ratios, and using $E(B-V)$ values provided by \citet{harris2010} listed in Table~\ref{tbl:gc_data}. The extinction-free Wesenheit magnitudes \citep{madore1982} in NIR filters were defined as in  \citet{bhardwaj2016a}: $W_{J,H}=H-1.83(J-H),~W_{J,K_s}=K-0.69(J-K_s),~W_{H,K_s}=K-1.92(H-K_s)$. 

We simultaneously solved for individual GC distances and derived PLZ relations for the global sample of 1310 RR Lyrae stars. In the case of RRc variables, we also added 52 RRc models with $0.25 < P < 0.45$~days \citep{marconi2015} to the calibrator sample because there are only 25 MW field RRc stars. These RRc models allowed us to break the degeneracy between the metallicity coefficient and the zero-points. Table~\ref{tbl:plzr} lists the results of the best-fitting linear regression formalism discussed in Section~\ref{sec:lfit}. Fig.~\ref{fig:plzrs} displays the PLZ/PWZ relations for the combined sample of RR Lyrae variables in GCs. The metallicity effect on RR Lyrae magnitudes is distinctly evident in the top panels showing that the metal poor stars are brighter. Tighter PLZ/PWZ relations are seen in the bottom panels where the metallicity term is taken into account, and the residuals do not exhibit any significant trend as a function of period or metallicity. These empirical metallicity coefficients are the most precisely determined in $JHK_s$ filters, and are in excellent agreement with the horizontal branch evolution and pulsation model predictions. There is a hint of increasing metallicity dependence at shorter wavelengths, but the differences are not statistically significant. We did not find any statistically significant changes in the PLZ relations for different metallicity scales and reddening values as discussed in the appendix ~\ref{app:test}.

\begin{deluxetable*}{rrrrrrrrrrrr}
\huge
\tablecaption{NIR period-luminosity-metallicity and period-Wesenheit-metallicity relations. \label{tbl:plzr}}
\tabletypesize{\footnotesize}
\tablewidth{0pt}
\tablehead{\colhead{\bf Band}  & \colhead{\bf Type} & \colhead{\bf a}  &  \colhead{\bf b}& \colhead{\bf c} & \colhead{\bf $\sigma$}   & \colhead{\bf $N_\textrm{f}$}   & \colhead{\bf $\chi^2_\textrm{dof}$} & \colhead{\bf $\Delta \mu$} & \colhead{\bf $\Delta\textrm{\bf a}$}  & \colhead{\bf $\Delta\textrm{\bf b}$}   & \colhead{\bf $\Delta\textrm{\bf c}$ } 
}
\startdata
          $J$&    RRL & $   -0.44\pm0.03    $ & $   -1.83\pm0.02    $ & $    0.20\pm0.02    $ &     0.05&  1049&     1.03& $    0.02\pm0.05    $ & $   -0.06$ & $   -0.07$ & $   -0.02$ \\
          $H$&    RRL & $   -0.74\pm0.02    $ & $   -2.29\pm0.02    $ & $    0.19\pm0.01    $ &     0.05&  1037&     0.98& $   -0.01\pm0.04    $ & $   -0.02$ & $    0.06$ & $   -0.01$ \\
        $K_s$&    RRL & $   -0.80\pm0.02    $ & $   -2.37\pm0.02    $ & $    0.18\pm0.01    $ &     0.05&  1077&     1.02& $   -0.01\pm0.03    $ & $   -0.02$ & $    0.12$ & $   -0.00$ \\
    $W_{J,H}$&    RRL & $   -1.20\pm0.02    $ & $   -2.96\pm0.02    $ & $    0.18\pm0.01    $ &     0.07&   989&     1.07& $   -0.05\pm0.04    $ & $    0.01$ & $    0.20$ & $    0.02$ \\
  $W_{J,K_s}$&    RRL & $   -1.06\pm0.03    $ & $   -2.73\pm0.02    $ & $    0.16\pm0.02    $ &     0.06&  1096&     0.99& $   -0.04\pm0.03    $ & $    0.01$ & $    0.23$ & $    0.02$ \\
  $W_{H,K_s}$&    RRL & $   -0.93\pm0.02    $ & $   -2.45\pm0.02    $ & $    0.16\pm0.01    $ &     0.07&  1033&     1.17& $   -0.01\pm0.04    $ & $   -0.01$ & $    0.13$ & $    0.02$ \\
\hline
          $J$&   RRab & $   -0.46\pm0.02    $ & $   -1.97\pm0.04    $ & $    0.20\pm0.01    $ &     0.05&   651&     0.83& $    0.03\pm0.05    $ & $   -0.05$ & $   -0.01$ & $    0.01$ \\
          $H$&   RRab & $   -0.74\pm0.02    $ & $   -2.28\pm0.03    $ & $    0.19\pm0.01    $ &     0.05&   667&     0.88& $   -0.01\pm0.04    $ & $   -0.02$ & $    0.04$ & $   -0.00$ \\
        $K_s$&   RRab & $   -0.81\pm0.03    $ & $   -2.39\pm0.03    $ & $    0.18\pm0.02    $ &     0.05&   709&     0.99& $   -0.02\pm0.04    $ & $   -0.01$ & $    0.12$ & $    0.00$ \\
    $W_{J,H}$&   RRab & $   -1.19\pm0.03    $ & $   -2.91\pm0.04    $ & $    0.18\pm0.02    $ &     0.07&   632&     0.88& $   -0.06\pm0.04    $ & $    0.02$ & $    0.24$ & $    0.02$ \\
  $W_{J,K_s}$&   RRab & $   -1.07\pm0.03    $ & $   -2.69\pm0.04    $ & $    0.16\pm0.02    $ &     0.06&   714&     0.84& $   -0.05\pm0.05    $ & $    0.04$ & $    0.20$ & $    0.03$ \\
  $W_{H,K_s}$&   RRab & $   -0.93\pm0.03    $ & $   -2.41\pm0.04    $ & $    0.16\pm0.01    $ &     0.07&   680&     0.99& $    0.00\pm0.06    $ & $    0.01$ & $    0.07$ & $    0.01$ \\
\hline
          $J$& RRc+RRd &    $-0.89\pm0.04$    & $   -2.14\pm0.05    $ & $    0.17\pm0.03  $  &     0.05&   437&     1.02& $    0.03\pm0.04    $ & $   -0.18$ & $   -0.32$ & $   -0.02$ \\
          $H$& RRc+RRd &    $-1.23\pm0.04$    & $   -2.51\pm0.05    $ & $    0.17\pm0.03  $  &     0.05&   428&     0.93& $    0.04\pm0.02    $ & $   -0.08$ & $   -0.19$ & $   -0.01$ \\
        $K_s$& RRc+RRd &    $-1.30\pm0.04$    & $   -2.60\pm0.05    $ & $    0.17\pm0.03  $  &     0.05&   435&     1.03& $    0.04\pm0.03    $ & $   -0.08$ & $   -0.12$ & $   -0.02$ \\
    $W_{J,H}$& RRc+RRd &    $-1.72\pm0.04$    & $   -2.92\pm0.06    $ & $    0.18\pm0.03  $  &     0.06&   392&     1.08& $    0.09\pm0.05    $ & $    0.00$ & $   -0.16$ & $   -0.01$ \\
  $W_{J,K_s}$& RRc+RRd &    $-1.55\pm0.04$    & $   -2.88\pm0.05    $ & $    0.18\pm0.03  $  &     0.05&   422&     1.07& $    0.06\pm0.03    $ & $   -0.03$ & $   -0.02$ & $   -0.02$ \\
  $W_{H,K_s}$& RRc+RRd &    $-1.37\pm0.04$    & $   -2.71\pm0.05    $ & $    0.18\pm0.03  $  &     0.06&   410&     1.22& $    0.07\pm0.04    $ & $   -0.11$ & $   -0.04$ & $   -0.03$ \\
  \hline
\enddata
\tablecomments{The zero-point ($a$), slope ($b$), metallicity coefficient ($c$), dispersion ($\sigma$), number of stars ($N_f$), and the chi-square per degree of freedom ($\chi^2_\textrm{dof}$) of the final PLR fits are tabulated. $\Delta\mu$ represents the mean and the standard deviation of residuals between distance moduli from this work and those from \citet{baumgardt2021}. The last three columns show the difference in the zero-points, the slopes, and the metallicity coefficients between predicted \citep{marconi2015} and empirical relations derived in this study. In the case of RRc+RRd stars only, the calibrator sample of MW field stars also includes 52 models from \citet{marconi2015} as mentioned in the text.}
\vspace{-05pt}
\end{deluxetable*}

Figure~\ref{fig:plzrs} also shows predicted PLZ/PWZ relations from \citet[][corresponding to {[Fe/H]}$~=-0.5,-1.5$ and $-2.5$ dex]{marconi2015} which suggest a good agreement with our empirical relations. The last three columns of Table~\ref{tbl:plzr} list the differences in the slopes, zero-points, and the metallicity coefficients between theoretical  \citep{marconi2015} and empirical PLZ/PWZ relations derived in this work. While the metallicity coefficients are very similar, there are some differences in the slopes and zero-points for some specific sample and/or filter combinations. In particular, the slope and zero-point of RRc stars in $J$-band are significantly shallower and fainter than the model predictions, respectively. 
We note that the error in the predicted slope for RRc stars in $J$-band is $0.08$~mag \citep{marconi2015}, and therefore, the difference of $-0.32$~mag in the slopes between predicted and empirical relations is just beyond $3\sigma$ of their combined uncertainties. Similarly, the largest difference in zero-point ($-0.18\pm0.06$ mag)  also occurs for RRc stars in the $J$-band, but it is within $\pm0.05$~mag for most sample and/or filter combinations. 

\subsection{Distance moduli to globular clusters}

Table~\ref{tbl:gc_data} also lists the average distance moduli to the GCs derived after solving equations~(\ref{eq:plzr_gc}) and (\ref{eq:plzr_mw}), for PLZ and PWZ relations for the sample of all 1310 RR Lyrae stars. Fig.~\ref{fig:delmu} in the appendix \ref{app:fig} shows no obvious trend in the difference between the distance moduli derived in this work and those from \citet{baumgardt2021} as a function of metallicity. While the residual offset in average distance moduli is statistically small for all RR Lyrae sample ($-0.026\pm0.025$~mag), the RR Lyrae distances to GCs are systematically smaller using empirical calibrations for all/RRab sample and larger using semi-empirical calibrations for RRc+RRd sample (see $\Delta\mu$ in Table~\ref{tbl:plzr}). 
\citet{bhardwaj2021} also found a similar result with absolute zero-points of $-0.76/-0.84$~mag in $K_s$ using empirical/theoretical calibrations. The empirical calibrations use {\it Gaia} distances taking into account the parallax zero-point offset from \citet{lindegren2021}. A fainter absolute zero-point of empirical PLZ/PWZ relation, which results in smaller distance moduli, suggests an over-correction in parallaxes because a smaller parallax zero-point offset than the \citet{lindegren2021} correction will result in a brighter zero-point for RR Lyrae stars \citep[see][]{bhardwaj2021}. A large scatter in the PLZ relations for MW field stars (Fig.~\ref{fig:calib}) precludes us from quantifying this small offset in parallaxes in the latest {\it Gaia} data. Comparison of distance moduli listed in Table~\ref{tbl:gc_data} also suggests that the difference is larger for metal-rich GCs while the metal-poor GCs show excellent agreement. This difference can be attributed to a small number of both RR Lyrae stars and GCs for [Fe/H]~$<-1.2$~dex (see Fig.~\ref{fig:hists}), which could affect the accuracy of the apparent zero-points of PLZ relations, and in turn distance moduli, for the relatively metal-rich GCs.

\subsection{Validation of empirical period-luminosity-metallicity relations}

To validate the accuracy of our empirically derived PLZ and PWZ relations, we determined a distance to the LMC using available NIR photometry in the literature. We use $\sim 18500$ RR Lyrae stars (13611 RRab, 4848 RRc) from \citet{cusano2021} with $JK_s$ photometry from the VMC survey together with their photometric metallicity and reddening estimates. Our empirical calibrations result in an average distance modulus to the LMC of $18.54\pm0.13$, $18.57\pm0.12$, and $18.60\pm0.15$ mag in $K_s$ band using the whole sample, RRab, and RRc+RRd stars, respectively. The quoted values and their uncertainties represent the peak and the width of the Gaussian distribution taking into account the uncertainties on the metallicity scale and on individual metallicities. When using theoretical relations of \citet{marconi2015}, we obtained a distance of $18.60\pm0.13$, and $18.59\pm0.15$ mag for all/RRab and RRc+RRd samples, respectively, which reflects the difference in the slopes and zero-points between empirical and theoretical calibrations. The difference in the LMC distance modulus with the percent-level precise value of 18.477$\pm$0.026~mag \citep{piet2019} is likely due to the differences in the calibrator and target PLZ relations, and due to uncertainties in photometric metallicities and reddening, and geometrical effects, which contribute to three times larger scatter in the PLZ relations for RR Lyrae stars in the LMC. 

\section{Conclusions}

We presented the results of a NIR monitoring program of GCs to quantify the metallicity dependence on RR Lyrae PLZ and PWZ relations. The largest sample of 964 cluster RR Lyrae stars with homogeneous $JHK_s$ mean magnitudes based on time-series data in 11 GCs of different mean metallicities was used in this analysis. When anchored using additional 346 MW field RR Lyrae stars with {\it Gaia} parallaxes, we obtained purely empirical PLZ and PWZ relations that agree well with the theoretical predictions based on horizontal branch evolution and pulsation models. The metallicity coefficient of NIR PLZ and PWZ relations is $\sim 0.2$ mag/dex, in excellent agreement with the predicted relations. We simultaneously obtained distance moduli to individual GCs and found those to be consistent with most recent and independent measurements in the literature based on {\it Gaia} data. While the metallicity coefficients and the slopes are firmly constrained, the absolute zero-points of PLZ/PWZ relations are sensitive, at the percent-level precision, to the uncertainties in parallaxes, parallax zero-point offset, lack of homogeneous time-series photometry and spectroscopic metallicities for MW field stars. Furthermore, any possible variations in the slopes of the underlying PLZ/PWZ relations in different stellar environments may also affect the zero-point, and this basic assumption of universality of the PLRs will be investigated in detail in a future study. Nevertheless, our accurate and precise empirical calibrations of PLZ/PWZ relations will be crucial for RR Lyrae based distance measurements and the calibration of the distance scale based on population II standard candles in the era of space telescopes and next-generation ground-based facilities operating mainly at infrared wavelengths.  

\acknowledgments
We thank the anonymous referee for the constructive referee report that helped improve the manuscript.
This project has received funding from the European Union’s Horizon 2020 research and innovation programme under the Marie Skłodowska-Curie grant agreement No. 886298. 
This research was supported in part by the Australian Research Council Centre of Excellence for All Sky Astrophysics in 3 Dimensions (ASTRO 3D), through project number CE170100013. 
HPS acknowledges grant 03(1428)/18/EMR-II from
Council of Scientific and Industrial Research (CSIR), India.
CCN thanks the funding from the National Science and Technology Council (Taiwan) under the contract 109-2112-M-008-014-MY3. Access to the CFHT was made possible by the Institute of Astronomy and Astrophysics, Academia Sinica. 

\facility{CFHT-WIRCam, Gemini FLAMINGOS-2}
\software{\texttt{IDL} \citep{landsman1993},
\texttt{Astropy} \citep{astropy2013}}\\

\vspace{200pt}
\appendix

\section{Impact of different metallicity scales and reddening values}
\label{app:test}

We tested for possible variations in the PLZ relations due to different metallicity scales and reddening values in the literature using the full sample of 1310 RR Lyrae stars. Our calibrator sample of 346 RR Lyrae stars was adopted from \citet{muraveva2018} which was based on the homogenized compilation of spectroscopic and photometric metallicities by \citet{dambis2013} in \citet{zinn1984} metallicity scale. \citet{crestani2021} published high resolution spectroscopic metallicities for 208 MW field RR Lyrae stars. There are 166 RR Lyrae stars in the \citet{crestani2021} sample after restricting that to the same metallicity range and 2MASS quality flag criteria as in Section~\ref{sec:data}. The coefficients of the PLZ relations derived using these 166 stars together with 964 GCs variables are statistically consistent with those listed in Table~\ref{tbl:plzr} within $1\sigma$ of their quoted uncertainties. This was expected considering a small median offset of $\Delta$[Fe/H]$~=0.09\pm0.02$~dex for 133 common stars in the two calibrator samples of MW field stars. 

In the case of mean metallicities of GCs, the two sets of independent estimates were obtained from the catalogs of \citet{harris2010} and \citet{dias2016}. The mean difference in the [Fe/H] values of 11 GCs is $0.03$~dex between the estimates from \citet{harris2010} and \citet{carretta2009}. This difference in the [Fe/H] values is $0.08$~dex between the measurements of \citet{dias2016} and \citet{carretta2009}. Given these small mean differences with our adopted metallicities, the coefficients of PLZ relations were found to be in agreement with the values listed in Table~\ref{tbl:plzr} within $0.5\sigma$ of their uncertainties, when using the aforementioned two independent sets of mean [Fe/H] values.

To investigate the impact of the adopted extinction law and reddening values, we obtained $E(B-V)$ values from the Baystar19 3D dust map \citep{green2019}. The Baystar19 map does not cover NGC 6441 or NGC 5139, for which $E(B-V)$ values were adopted from \citet{schlegel1998}. The $E(B-V)$ values from these maps are in the same units and the corresponding absorption ratios in $JHK_s$ were taken from \citet[][Table 1]{green2019}. After applying these independent reddening corrections, we derive $JHK_s$ band PLZ relations which exhibit the same slope as listed in Table~\ref{tbl:plzr}. The metallicity coefficients increase by 0.01~dex in all three NIR filters and the zero-points are fainter by 0.04/0.02/0.02 mag in $J/H/K_s$ bands. These differences in the $J$ and $HK_s$ band zero-points are within $1.4\sigma$ and $1\sigma$ of their associated uncertainties, respectively.  

\section{Difference in the distance moduli to globular clusters}
\label{app:fig}

\begin{figure}
\centering
\includegraphics[width=0.48\textwidth]{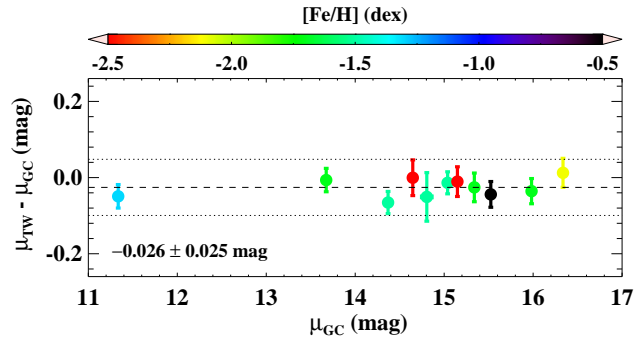}\\
\caption{The difference between the distance moduli ($\mu_\textrm{TW}$) resulting from equation~(\ref{eq:plzr_gc}) and those from \citet[][$\mu_\textrm{GC}$]{baumgardt2021}, as listed in Table~\ref{tbl:gc_data}. The mean and the standard deviation of residuals are quoted at the bottom left. The dashed and dotted lines represent the mean and three times the standard deviation of the residuals, respectively.}
\label{fig:delmu}
\end{figure}

Fig.~\ref{fig:delmu} displays the difference in the distance moduli listed in the last two columns of Table~\ref{tbl:gc_data}.

\bibliographystyle{aasjournal}
\bibliography{mybib_final.bib}

\end{document}